\documentclass[12pt]{article}

\def\theequation{\arabic{equation}}

  \def\theequation{\thesection.\arabic{equation}}
  \def\theequation{\arabic{equation}}

\newsavebox{\savepar}

\newtheorem{remark}{Remark}[section]
\newtheorem{remarks}{Remarks}[section]
\newtheorem{example}{Example}[section]

\newtheorem{theorem}{Theorem}[section]

\newtheorem{lemma}{Lemma}[section]

\def\appendix{\par
   \setcounter{section}{0}
   \setcounter{equation}{0}
   \def\@chapapp{APPENDIX}
   \def\thesection{\Alph{section}}
   \def\theequation{A.\arabic{equation}}
   \def\section{
   \setcounter{equation}{0}
   \refstepcounter{section}
   \@startsection {section}{A}{0pt}{-3.5ex \@plus -1ex \@minus -.2ex}
                  {0.3ex \@plus.2ex}{\normalsize\sffamily\bfseries} }}

\def\eop{\hbox{{\vrule height7pt width3pt depth0pt}}}

\makeatletter
\newcommand{\least}{\let\CS=\@currsize\renewcommand{\baselinestretch}{1}\tiny\CS}

\newcommand{\oneandahalfspacing}{\let\CS=\@currsize\renewcommand{\baselinestretch}{1.2}\tiny\CS}
\newcommand{\doublespacing}{\let\CS=\@currsize\renewcommand{\baselinestretch}{2.5}\tiny\CS}

\usepackage{amsmath}
\usepackage{amsfonts}
\usepackage{euscript}
\usepackage{oldgerm}
\usepackage{rotating}

  \renewcommand{\baselinestretch}{1.3}

\makeindex

\begin{document}

\newcommand{\namelistlabel}[1]{\mbox{#1}\hfil}
\newenvironment{namelist}[1]{%
\begin{list}{}
{
\let\makelabel\namelistlabel
\settowidth{\labelwidth}{#1}
\setlength{\leftmargin}{1.1\labelwidth}
}
}{%
\end{list}}

\numberwithin{equation}{section}
\newcommand{\be}{\begin{equation}}
\newcommand{\ee}{\end{equation}}
\newcommand{\dist}{{\rm\,dist}}
\newcommand{\sspan}{{\rm\,span}}
\newcommand{\re}{{\rm Re\,}}
\newcommand{\im}{{\rm Im\,}}
\newcommand{\sgn}{{\rm sgn\,}}
\newcommand{\beano}{\begin{eqnarray*}}
\newcommand{\eeano}{\end{eqnarray*}}
\newcommand{\bea}{\begin{eqnarray}}
\newcommand{\eea}{\end{eqnarray}}

\newcommand{\ba}{\begin{array}}
\newcommand{\ea}{\end{array}}
\newcommand{\hone}{\mbox{\hspace{1em}}}
\newcommand{\hon}{\mbox{\hspace{1em}}}
\newcommand{\htwo}{\mbox{\hspace{2em}}}
\newcommand{\hthree}{\mbox{\hspace{3em}}}
\newcommand{\hfour}{\mbox{\hspace{4em}}}
\newcommand{\von}{\vskip 1ex}
\newcommand{\vone}{\vskip 2ex}
\newcommand{\vtwo}{\vskip 4ex}
\newcommand{\vthree}{\vskip 6ex}
\newcommand{\vfour}{\vspace*{8ex}}
\newcommand{\norm}{\|\;\;\|}
\newcommand{\integ}[4]{\int_{#1}^{#2}\,{#3}\,d{#4}}
\newcommand{\inp}[2]{\langle {#1} ,\,{#2} \rangle}
\newcommand{\vspan}[1]{{{\rm\,span}\{ #1 \}}}
\newcommand{\R} {{\mathbb{R}}}

\newcommand{\B} {{\mathbb{B}}}
\newcommand{\C} {{\mathbb{C}}}
\newcommand{\N} {{\mathbb{N}}}
\newcommand{\Q} {{\mathbb{Q}}}
\newcommand{\LL} {{\mathbb{L}}}
\newcommand{\Z} {{\mathbb{Z}}}

\newcommand{\BB} {{\mathcal{B}}}
\newcommand{\dm}[1]{ {\displaystyle{#1} } }
\def \stackt{{\stackrel{.}{.\;.}\;\;}}
\def \stackb{{\stackrel{.\;.}{.}\;\;}}
\def \olu{\overline{u}}
\def \olv{\overline{v}}
\def \olx{\overline{x}}
\def \olp{\overline{\partial}}
\def\diag{{\;{\rm diag } \; }}
\thispagestyle{empty}
\bibliographystyle{}

\begin{center}
{\Large \bf The Unbiasedness Approach to Linear Regression Models}\\
\end{center}

\vtwo
\begin{center}
{\large \bf  P. Vellaisamy \\
{\it Department of Mathematics, Indian Institute of Technology Bombay, \\
Powai, Mumbai-400076, India.}}
\end{center}

\vone

\abstract{ The linear regression models are widely used statistical techniques in numerous practical applications. The standard regression model requires several assumptions about the regressors and the error term. The regression parameters are estimated using the least-squares method. In this paper,
we consider the regression model with arbitrary regressors  and without the error term. An explicit expression for the
regression parameters vector is obtained. The unbiasedness approach is used to estimate the regression parameters and its various properties are investigated. It is shown that the resulting unbiased estimator equals the least-squares estimator for the fixed design model. The analysis of residuals and the regression sum of squares can be carried out in a natural way. The unbiased estimator of the dispersion matrix of the
unbiased estimator is also obtained. Applications to $AR(p)$ model and numerical examples are also discussed.}
\vone

\section{Introduction}
The linear regression model is a commonly used statistical technique in practical applications (Quenonuille (1950)), because of
its simplicity and its realistic nature for modeling several phenomena. For an extensive treatment of this topic, one may refer to  Draper and Smith (2003) and Chatterji and Hadi (2003). In the standard multiple linear regression model $Y=\beta_1X_1+\ldots,\beta_p X_p +\epsilon$, it is  generally assumed that the predictor variables  $X_1, \ldots, X_p$ are  constant (non-stochastic) and also the error term
$\epsilon$ is independent of predictor variables (see, for example, Yan and Su (2009)), especially for the estimation of 
regression coefficients. This means that the values of $X= (X_1, \ldots, X_p)$
are controlled by an experimenter and when the experiment is repeated the values of $X$ does not change, but $Y$ changes as $\epsilon$ changes. These assumptions are not realistic in practice and in most experimental
setups and especially in econometric situations where the changes in $Y$ occurs mainly due to the changes in $X$.
Even under the stochastic regressors models, it is commonly assumed that $X_i$'s are independent of the $\epsilon$. It is known that when one of the $X_i$ is correlated with $\epsilon_i$,
the ordinary least-squares (OLS) estimator   becomes biased and  inconsistent as well.  To overcome such
difficulties, an instrumental variable (IV) $Z$, which is highly correlated with $X_i$ and is independent of
$\epsilon_i$, is discussed in the literature. But, finding a suitable IV is also an issue. Also, OLS estimator
has smaller variance with IV estimator, though it may have bias  larger that that of IV estimator.

\vone
\noindent In this paper, we consider the linear regression model with little assumptions about the predictors
and treat them as random. We do not explicitly introduce the error term and thus avoid not only the associated probabilistic assumptions, but also the related issues mentioned above. Besides, we do not use the least-squares
method for estimating the regression coefficients which requires vector differentiation and related minimization problems. In Section 2, we consider the linear regression model with stochastic regressors and obtain first an explicit relationship between the regression coefficients and the characteristics of the distribution of $(Y, X_1, \ldots, X_p)$.  In Section 3, we obtain the unbiased estimators, using the relationship between regression coefficients. Our approach is semi-parametric in nature and does not assume any specific distribution, such as multivariate normal, which is commonly used in literature. Note that the  unbiased estimator coincides with OLS estimator, as they are conditionally unbiased.
 Several known properties of the unbiased estimators are established using our approach. The dispersion matrix of the estimator is derived and also its unbiased estimator is also observed. The mean regression sum of squares and also the mean residual sum of
squares are derived in a natural way. Using these results,
a predictor of $E(Y)$ based on a new observation is also pointed out.  The method is then extended to autoregressive $AR(p)$ model also. Finally, we  discuss two examples, one based on simulated data and other one based on a real-life data, to show that that our approach yields the same least-squares estimates for the  $AR (3) $ model.

\section {Multiple Linear Regression Model}

Let $Y$ be the response or dependent variable, and $X_1, X_2, \ldots, X_p$ be $p$ arbitrary explanatory (predictor variables) which are  not necessarily independent. Consider the class $\cal{C}$ of linear regression model defined by
         \begin{eqnarray} \label{eqn2.1}
    E(Y|X_1, \dots , X_p) &=& b_0 + b_1X_1+ \ldots +b_pX_p \nonumber \\
                       &=&b_0+X_{(p)}b_{(p)}, (say) ,
     \end{eqnarray}
where $X_{(p)}= (X_1,\ldots,X_p)$ is the row-vector of  predictor variables, and  $b^{}_{(p)}= (b_1,\ldots, b_p)^{t}$ is the column-vector of regression coefficients.
Note that model (\ref{eqn2.1}) is rather general and includes the class $\cal{C}_{\epsilon}$ of the usual linear regression models, defined by
\begin{equation} \label{neqn2.1}
    Y = b_0 + b_1X_1+ \ldots +b_pX_p + \epsilon,  
 \end{equation}                      
where $\epsilon$ is the error term independent of the $X_i$'s.
  For example, let
 $Z\perp (X_1,X_2)$, $E(Z)=0$ and consider  the model
\begin{equation} \label{eqn2.1n}
Y=b_1X_1+b_2X_2+ b_3|X_1-X_2|Z. 
\end{equation}
Then
$E(Y|X_1,X_2)=b_1X_1+b_2X_2 $ which belongs to $\cal{C}$, but does not belong to $\cal{C}_{\epsilon}$. Thus, $\cal{C}_{\epsilon} \subset \cal{C}.$ As another example, consider the autoregressive $AR(p)$ process defined by
\begin{equation} \label{neqn2.1n}
X_t=\phi_0+ \phi_1X_{t-1}+\phi_2X_{t-2}+ \ldots+ \phi_p X_{t-p} + \epsilon, 
\end{equation}
where the error term is independent of the $X_i, ~ (t-p) \leq i \leq t.$ Then
\begin{equation*}
    E(X_t|X_1, \dots , X_{t-1}) = \phi_0+ \phi_1X_{t-1}+\phi_2X_{t-2}+ \ldots+ \phi_p X_{t-p}
 \end{equation*}                      
which is again of the from given in  (\ref{eqn2.1}). Note here the $X_i$ are not independent. Thus, $AR(p)$ process belongs to $\cal{C}_{\epsilon}$  and hence also to $\cal{C}$.

\noindent Our aim here is to consider the linear regression model defined in \eqref{eqn2.1} and the estimation of
regression coefficients in a natural and simple way, with hardly a few assumptions on the model. For that purpose, we  require the relationship among regression coefficients.

\subsection {Relationships Among Regression Coefficients}
\noindent Let $Y_0=(Y-b_0)$ so that from \eqref{eqn2.1}
\begin{equation} \label{eqn2.2}
E(Y_0|X_{(p)})=X_{(p)}b_{(p)};~~~~
E(Y_0)= E(X_{(p)})b_{(p)},
\end{equation}
where $E(X_{(p)})=(EX_{1},\ldots,EX_{p}) $ is the row vector of the means of $X_{i}$'s.
\noindent  First our aim is to obtain an explicit representation for the regression coefficients.
From \eqref{eqn2.2}, we get
\begin{equation} \label{eqn2.3}
E(Y_0X_1|X_{(p)})=X_{(p)}X_1b_{(p)}; ~~~~
E(Y_0X_1)= E (X_{(p)}X_1) b_{(p)}. \end{equation}

\noindent Hence, from \eqref{eqn2.2} and \eqref{eqn2.3}
\begin{eqnarray}
Cov(Y_0,X_1) = E(X_{(p)}X_1)b_{(p)} -E(X_{(p)})E(X_1)b_{(p)}= Cov(X_{(p)},X_1)b_{(p)},          
\end{eqnarray}
where
\begin{equation*}
Cov(X_{(p)},X_1)=\left(Cov(X_1,X_1),Cov(X_2,X_1),\ldots,Cov(X_{(p)},X_1)\right)
\end{equation*}
is the row-vector of covariances. Similarly, $Cov(Y_0,X_i) = Cov(X_{(p)},X_i)b_{(p)}, ~ 2 \leq i \leq p.$

\noindent Since $Cov(Y_0,X_{i})=Cov(Y,X_{i}),$
we can represent the above observations in a matrix form as,
\begin{eqnarray} \label{eqn2.7}
\begin{pmatrix}                  
Cov(Y, X_{1})\\                  
Cov(Y, X_{2})\\                
\vdots \\                          
Cov(Y, X_{p})
\end{pmatrix}                                          
=
\begin{pmatrix}
C_{11}, C_{21}, \ldots , C_{p1}\\
C_{12}, C_{22}, \ldots , C_{p2}\\
\vdots\\
C_{1p}, C_{2p}, \ldots , C_{pp}
\end{pmatrix}
\begin{pmatrix}
b_{1}\\
b_{2}\\
\vdots\\
b_{p}
\end{pmatrix},
\end{eqnarray}
where $ C_{ij} = Cov(X_{i}, X_{j}), ~\mbox{for} ~1\le i, j \le p. $
Let now
\begin{eqnarray} \label{neqn2.7}
C^{t}_{yx} =(C_{y1} , C_{y2} , \ldots , C_{yp})
          =(Cov(Y,X_1), \ldots, Cov(Y,X_{p}); ~~
   C_{xx} = (C_{ij}).
\end{eqnarray}
Then equation {\eqref{eqn2.7}} can be written compactly as
\begin{eqnarray} \label{eqn2.8}
C_{yx} = C_{xx}b_{(p)},
\end{eqnarray}
which gives the relationship between regression coefficients and the associated covariances.
Also from  {\eqref{eqn2.2}}, we obtain
\begin{eqnarray}  \label{eqn2.9}
b_{0}&=& E(Y)-E(X_{(p)})b_{(p)}.
\end{eqnarray}  
\noindent Thus, we have proved the following result.
\begin{theorem}\label{thm2.1}
 Let $E(Y|X_1,\ldots X_p)=b_0+X_{(p)}b_{(p)}$ be the multiple linear regression model defined as in \eqref{eqn2.1}.
Let $C_{yx}^t=( C_{y1}, C_{y2},\ldots, C_{yp})$, ~with $C_{yj}=Cov(Y,X_j)$,
and $C_{xx}=(C_{ij})$ with $C_{ij}=Cov(X_i,X_j),  1 \le i , j \le p$.
Then the regression coefficients $b_0$ and $b_{(p)}$ satisfy
\begin{eqnarray} \label{eqn2.10n}
C_{yx}=C_{xx}b_{(p)}, ~~\mbox{and}
\end{eqnarray}
\begin{eqnarray} \label{eqn2.11n}
b_0=\mu_y-\mu_{(p)}b_{(p)},
\end{eqnarray}
where $\mu_y=E(Y)$ and $\mu_{(p)}=(EX_1,\cdots,EX_p) =(\mu_{1},\ldots,\mu_{p}).$
\end{theorem}

\noindent When $C_{xx}$ is  non-singular, we have a unique representation as
\begin{equation} \label{eqn2.12n}
  b_{(p)}= C_{xx}^{-1} C_{yx}.
\end{equation}
If $C_{xx}$ is singular, a g-inverse $C_{xx}^{-}$ may be used to obtain a representation for $b_{(p)}$.
\noindent That is, using Jordan decomposition, let
$C_{xx}=PDP^t$,
where $P$ is an orthogonal matrix, and \\
$D=diag[\lambda_1, \ldots, \lambda_p]$ is a diagonal matrix. Then a g-inverse of $C_{xx}$ is
$\overline{C}_{xx}=P \overline{D}P,$
where $\overline{D}=diag[\overline{\lambda}_1, \ldots, \overline{\lambda}_p]$, and $ \overline{\lambda}_i= \lambda^{-1}_i$ if $\lambda_i>0$, and is zero if $ \lambda_i=0$.
\vone

\begin{remarks} {\em
 Observe that when $p=1$ and  $ E(Y|X_1)= b_0 + b_1 X_1$, the regression coefficient $b_1$ is given by 
 \begin{equation*}
 b_1= \frac{C_{y1}} {C_{11}}= \frac{Cov (Y, X_1)} {Var (X_1)},
\end{equation*}
 a known result.

\noindent  Similarly, from \eqref{eqn2.12n}, the regression coefficients for the case $p=2$ are given by

\begin{equation}\label{eqn2.13n}
                 b_1= \frac{C_{22}C_{y1}-C_{12}C_{y2}} {C_{11}C_{22}- C_{12}^2};~~ b_2= \frac{C_{11}C_{y2}-C_{12}C_{y1}} {C_{11}C_{22}- C_{12}^2}
\end{equation}
and the intercept $b_0$ can be evaluated using \eqref{eqn2.11n} and \eqref{eqn2.13n}.}
\end{remarks}

\begin{remarks}  {\em The following properties of the model in \eqref{eqn2.1} follow now easily. \\
(i) It is known that $f(X)=E(Y|X)$ minimizes $E(Y-f(X))^2$. This shows that  $E(Y|X)=X_{(p)}b_{(p)}$ minimizes the mean squared error when $X_{(p)}b_{(p)}$ is treated as an approximation to $Y$.\\

\noindent (ii) Also, when $C_{xx}>0$ (positive definite)
\begin{eqnarray}\label{eqn1}
Cov(Y-X_{(p)}b_{(p)}, X_{(p)})&=& Cov(Y, X_{(p)})-Cov(X_{(p)}b_{(p)}, X_{(p)})\nonumber  \\
&=&C^{t}_{yx}-b^{t}_{(p)} Cov(X_{(p)}, X_{(p)})\nonumber \\
&=&C^{t}_{yx}-b^{t}_{(p)}C_{xx} \nonumber\\
&=&C^{t}_{yx}-C^{t}_{yx}C^{-1}_{xx}C_{xx}\nonumber \\
&=&0.
\end{eqnarray}

\noindent Indeed, we have from \eqref{eqn1},
\begin{eqnarray}
Cov(Y-X_{(p)}b_{(p)}, X_j)=0, \forall ~~i \leq j \leq p \nonumber\\
\Rightarrow Cov(Y-X_{(p)}b_{(p)}, X_{(p)}d_{(q)})=0
\end{eqnarray}
for any vector $d_{(q)}=(d_1, \ldots, d_q).$
In particular, we have
\begin{eqnarray} \label{eqn3}
Cov(Y-X_{(p)}b_{(p)}, X_{(p)}b_{(p)})=0.
\end{eqnarray}
Also,
\begin{eqnarray*}
Var(Y)&=&Var(Y-X_{(p)}b_{(p)}+ X_{(p)}b_{(p)})= Var(Y-X_{(p)}b_{(p)})+Var(X_{(p)}b_{(p)}),
\end{eqnarray*}
because of \eqref{eqn3}.}
\end{remarks}

\section {Estimation of Regression Coefficients}
We  start with a simple known result.
Let $(Y,X)$ be a bivariate random vector with with finite second moments and $ \sigma_{yx}=Cov(Y,X).$ 
Suppose $(Y,X_1),\ldots,(Y_n, X_n)$ is a random sample on the bivariate vector $(Y,X)$. Then it is well
known that
\begin{equation} \label{eqn3.1}
S_{yx}=\frac{1}{n-1}\sum_{i=1}^n(Y_{i}- \overline{Y})(X_{i}-\overline{X}_{i})
\end{equation}
is an unbiased estimator of $\sigma_{yx}$.

\noindent Consider now the estimation of regression coefficients $b_0$ and $b_{(p)}$ .
Suppose we have a random sample $(Y_{j},X_{1j},\ldots,X_{pj}),1 \le i \le n,$ of size $n$ on $(Y,X_1,X_2\ldots,X_p).$ Let
\begin{eqnarray} \label{eqn3.4n}
\mathbf{Y} =
\begin{pmatrix}
Y_{1}\\
\vdots\\
Y_{j}\\
\vdots\\
Y_{n}
\end{pmatrix}
;&&
\mathbf{X} =
\begin{pmatrix}
X_{(p)1}\\
\vdots\\
X_{(p)j}\\
\vdots\\
X_{(p)n}
\end{pmatrix}
=
\begin{pmatrix}
X_{11}~ X_{21} \ldots X_{i1} \ldots X_{p1}\\
\vdots\\
X_{1j}~ X_{2j} \ldots,X_{ij} \ldots X_{pj}\\
\vdots\\
X_{1n}~ X_{2n} \ldots X_{in},\ldots X_{pn}
\end{pmatrix},
\end{eqnarray}
where $X_{(p)j}=(X_{1j},X_{2j}, \ldots, X_{pj})$   denotes the row-vector of $j$-th observation on
predictor variables and $X_{ij}$ denotes the $j$-th observation on  $X_{i}, 1 \le i \le p, 1 \le j \le n. $
Then, from \eqref{eqn2.1},
\begin{equation} \label{eqn3.5n}
 E(Y_j|\mathbf{X})) =E(Y_j|X_{(p)j}) =b_0 + X_{(p)j}b_{(p)}, ~ 1 \leq j \leq n.
\end{equation}
\noindent Let $ \overline{Y}=\sum_{j=1}^n{Y_{j}}/n,$ and $ \overline{X_{i}}=\sum_{j=1}^n{X_{ij}}/n, 1 \le i \le p$ be the sample means.  Also, let
\begin{eqnarray}\label{eqn3.2correct}
S_{yi}&=&\frac{1}{n-1}\sum_{j=1}^n(Y_{j}- \overline{Y})(X_{ij}-\overline{X}_{i}) =\frac{1}{n-1}\sum_{j=1}^n(X_{ij}-\overline{X}_{i})Y_{j}
\end{eqnarray}
be the sample covariance between $Y$ and $X_i$ and
\begin{equation}\label{eqn3.3correct}
S_{lm} =\frac{1}{n-1}\sum_{j=1}^n(X_{lj}-\overline{X}_{l})(X_{mj}-\overline{X}_{m})\\
\end{equation}
be the sample covariances between $X_l$ and $X_m, 1 \le l,m \le p.$
Then by  \eqref{eqn3.1}, we have
\begin{equation}\label{eqn3.41correct}
E(S_{yi})= C_{yi}; ~~ E(S_{lm})= C_{lm},
\end{equation}
where $C_{yi}= Cov(Y, X_i)$ and $C_{lm}= Cov(X_l, X_m), 1 \leq l, m \leq n$.

\noindent Let $ S_{yx}^{t}=(S_{y1},S_{y2},\ldots,S_{yp}) $ be the row-vector and $ S_{xx}=(S_{ij})$ be the $(p\times p)$ matrix of sample covariances  of explanatory variables $X_1,X_2,\ldots,X_p.$
Then from \eqref{eqn3.41correct},
\begin{equation} \label{eqn3.5correct}
E(S_{yx})=C_{yx}; ~~  E(S_{xx})=C_{xx},
\end{equation}
where $ C_{xx}=(C_{ij})$ defined in \eqref{eqn2.7}.

\noindent Using Theorem \ref{thm2.1}, an estimator $\hat{b}_{(p)}$ of the regression coefficient vector $b_{(p)}$ can be obtained by solving
\begin{equation}\label{eqn3.6correct}
 S_{xx}\hat{b}_{(p)}=S_{yx},
\end{equation}
which is obtained by  replacing the unbiased estimator of $C_{yx}$ and $C_{xx}$ in \eqref{eqn2.10n}.\\
When $S_{xx}$ is nonsingular, {\it a.e}, an explicit and unique estimator of $b_{(p)}$ is
\begin{equation}\label{eqn3.7correct}
\hat{b}_{(p)}=S_{xx}^{-1}S_{yx}.\end{equation}
\noindent Similarly, an estimator of $b_0$ can be obtained from \eqref{eqn2.11n} as,
\begin{eqnarray}\label{eqn3.8correct}
\hat{b}_{0} &=& \hat{\mu}_{y}-\widehat{\mu}_{(p)}\widehat{b}_{(p)}=
\overline{Y}-  \overline{X}_{(p)} \widehat{b}_{(p)},
\end{eqnarray}where $\overline{X}_{(p)}=(\overline{X}_{1},\overline{X}_{2},\ldots,\overline {X}_{p}).$

\noindent For completeness, we next summarize the results discussed above.

\begin{theorem}\label{thm3.1}
Let $E(Y|X_{(p)})=b_0+X_{(p)}b_{(p)}$ and  $(Y_j,X_{1j},\ldots,X_{pj}), 1 \le j \le n,$ be a random sample of size $n$ on $(Y,X_{1},\ldots,X_{p})$. Let
$ S_{yx}=(S_{y1},S_{y2},\ldots,S_{yp})^{t}$be the column vector of sample covariances between $Y$ and $X_{i}$'s, and $S_{xx}=(S_{ij})$ be the sample covariance matrix of $X_{i}$'s. Let $S_{xx}$ be non-singular {\it a.e.} Then the natural estimators of $b_{(p)}$ and $b_{0}$ are
\begin{equation} \label{eqn3.9}\hat{b}_{(p)}=S_{xx}^{-1}S_{yx}, ~~~\mbox{and}~~ \end{equation}
\begin{equation} \label{eqn3.10}
\widehat{b}_{0} =\overline{Y}-  \overline{X}_{(p)} \widehat{b}_{(p)},\end{equation}
where $\overline{Y}$ and $\overline{X}_{(p)}=(\overline{X}_1,\ldots, \overline{X}_{p})$ denote the sample means.
\end{theorem}

\begin{remark} {\em
\noindent (i) When $p=1$ and $b_1=b,$ we have $\widehat{b}=S_{yx}/S_{xx},$ and $\widehat{b_0} = \overline{Y}-\overline{X}~\widehat{b}$,
which coincide with the classical least-squares estimators for the simple linear regression model.

\noindent (ii) From \eqref{eqn3.6correct}, $\widehat{b}_{(p)}$ satisfies in general $S_{xx}\widehat{b}_{(p)}=S_{yx}$. It can be shown that $S_{yx} \stackrel{p}\rightarrow C_{yx}$, as $ n \rightarrow \infty$, and $S_{xx} \stackrel{p}\rightarrow C_{xx}$, as a special case. Thus, by Slutsky's theorem, $ \widehat{b}_{(p)} \stackrel{p}\rightarrow b_{(p)}$, a solution of the defining equation  \eqref{eqn2.10n}. Hence, $\widehat{b}_{(p)}$ is a consistent estimator for $b_{(p)}$.

\noindent (iii) Note the natural estimators $ \widehat{b}_{(p)}$ and $\widehat{b}_{0}$ are obtained by substituting the unbiased estimator of $ C_{yx}, C_{xx}  $ and $\mu_{y}$ and $\mu_{(p)}$  in the defining equations \eqref{eqn2.10n} and \eqref{eqn2.11n}.

\noindent The next result shows that the natural estimators,  given in Theorem \ref{thm3.1}, are indeed unbiased.}
\end{remark}

\noindent Let henceforth $ \mathbf{X}= \mathbf{X}_{(n,p)}=(X_{1(n)}^t,\ldots,X_{p(n)}^t)=(X_{ij})$ be the matrix of observations (see \eqref{eqn3.4n}), where $X_{i(n)}= (X_{i1}, \ldots,
X_{in})$ denotes the row-vector of $n$ observations on $X_i$.

\begin{theorem}\label{thm3.2}
Let $S_{xx}$ be non-singular a.e. Then the estimators $\widehat{b}_{(p)}$ and $\widehat{b}_{0}$ are both conditionally unbiased and   hence are unbiased.
\end{theorem}
\noindent {\bf Proof.}
First note that, for $1 \le i \le p ,$
\begin{eqnarray*}
E(S_{yi}| \mathbf{X})&=& \frac{1}{(n-1)}\sum_{j=1}^nE\left((X_{ij}-\overline{X}_{i})Y_{j}| \mathbf{X}\right)\\
&=& \frac{1}{(n-1)}\sum_{j=1}^n(X_{ij}-\overline{X}_{i})E(Y_{j}| \mathbf{X})\\
&=& \frac{1}{(n-1)}\sum_{j=1}^n (X_{ij}-\overline{X}_{i})\left(b_{0}+\sum_{r=1}^pX_{rj}b_{r}\right) ~~ \mbox {(from \eqref{eqn3.5n})}\\
&=& \frac{1}{(n-1)}\sum_{j=1}^n(X_{ij}-\overline{X}_{i})\left(\sum_{r=1}^pX_{rj}b_{r}\right)\\
&=& \sum_{r=1}^p \left(\frac{1}{(n-1)}\sum_{j=1}^n(X_{ij}-\overline{X}_{i})X_{rj}\right)b_{r} ~~ \mbox{(interchanging the order)}\\
&=& \sum_{r=1}^p(S_{ir})b_{r}\nonumber\\
&=& S_{i(p)}b_{(p)},
\end{eqnarray*}
where $ S_{i(p)}=(S_{i1},\ldots,S_{ip})$.
Hence,
\begin{equation} \label{eqn3.16n}
E(S_{yx}| \mathbf{X})=S_{xx}b_{(p)} \Longrightarrow E(S_{xx}^{-1}S_{yx}| \mathbf{X})=E(\hat{b}_{(p)}| \mathbf{X}) =b_{(p)},
\end{equation}
showing that $\hat{b}_{(p)}$ is conditionally unbiased.

\noindent Similarly,
\begin{eqnarray} \label{neqn3.18}
E(\widehat{b}_{0}| \mathbf{X})&=&E(\overline{Y}| \mathbf{X})-\overline{X}_{(p)}E(\hat{b}_{(p)}| \mathbf{X}) \nonumber \\
&=&(b_0+b_1\overline{X}_1+\ldots+b_p\overline{X}_p| \mathbf{X})-\overline{X}_{(p)}b_{(p)} \nonumber ~~\mbox{ (using \eqref{eqn3.16n})}~~\\
&=&b_0+\overline{X}_{(p)}b_{(p)}-\overline{X}_{(p)}b_{(p)} \nonumber\\
&=& b_0,
\end{eqnarray}
and so $ \widehat{b}_{0}$ is also conditionally unbiased.

\noindent Hence, it follows, from \eqref{eqn3.16n} and  \eqref{neqn3.18}, that
\begin{eqnarray*}
 E(\hat{b}_{(p)})=b_{(p)};~~~ E(\hat{b}_0)=b_0,
\end{eqnarray*}
proving the result. \hfill{$\Box$}

\begin{remark} {\em (i)
Let now  $S_{xx} \geq 0~ a.e.$ (positive semidefinite) and $p>1.$  Then,
the estimator $\widehat{b}_{(p)}$ satisfies
\begin{eqnarray*}
S_{xx}\widehat{b}_{(p)}&=&S_{yx}, ~ \mbox{for  all}~(Y,\mathbf{X})\\
\Longrightarrow E(S_{xx} \widehat{b}_{(p)}|\mathbf{X})&=&E(S_{yx}|\mathbf{X}), ~ \mbox{for all}~ \mathbf{X}\\
\Longrightarrow S_{xx} \left(E(\widehat{b}_{(p)}|\mathbf{X})-b_{(p)}\right)&=&0, ~ \mbox{for  all}~\mathbf{X},
\end{eqnarray*}
using \eqref{eqn3.16n}.  Since $S_{xx} \geq 0 ~ a.e, $  we have
$ E(\widehat{b}_{(p)}|\mathbf{X}) = b_{(p)} ~ a.e $
and so $E(\widehat{b}_{(p)}) = b_{(p)}$, showing that $ \widehat{b}_{(p)}$ is unbiased in this case also.}
\end{remark}

\vone \noindent It is known that if $E(\widehat{\theta}_{i})=\theta_{i}, i=1,2,$ then
$E({\widehat{\theta}_{2}} /{\widehat{\theta}_{1}})=({\theta_{2}}/{\theta_{1}})$
may not hold in general. However, it holds in the linear regression case, in view of the following general result on a property of unbiased estimators. Let ${\cal{L}}(X)$ denote the distribution of $X.$

\begin{lemma}\label{3.3}
Let  $\theta_{1}$ be a characteristic of ${\cal{L}}(X)$, $\theta_{2}$ be a characteristic of ${\cal L}(Y,X)$  and $\theta= ({\theta_2}/{\theta_1})$.~ Let  $\widehat{\theta}_{1}=\widehat{\theta}_{1}(X)$ and $\widehat{\theta}_{2}=\widehat{\theta}_{2}(Y,X)$ be respectively estimators of $\theta_{1}$ and $\theta_{2}$ such that  $\widehat{\theta}_{1} >0 ~a.e$ and $E(\widehat{\theta}_{2}|X)=\theta \widehat{\theta}_{1}~a.e.$
Then $\widehat{\theta}= ({\widehat{\theta}_{2}}/{\widehat{\theta}_{1}})$ is also unbiased for $\theta$.
\end{lemma}
\noindent {\bf Proof.}
Since
\begin{equation}
E(\hat{\theta})=E(E(\hat{\theta}|X))= E \left(E\left(\frac{\widehat{\theta}_{2}}{\widehat{\theta}_{1}}|X \right)\right)
=E\left(\frac{1}{\widehat{\theta}_1}E(\widehat{\theta}_2|X)\right)
=E\left(\frac{1}{\widehat{\theta}_1}\theta\widehat{\theta}_{1}\right)=\theta,
\end{equation}
the result follows.  \hfill{$\Box$}

\begin{remark} {\em 
(i) Note in Lemma \ref{3.3}, $\widehat{\theta}_{1}$ is not necessarily unbiased. If in addition it is unbiased,
then we have
\begin{equation*}
E\left(\frac{\widehat{\theta}_{2}}{\widehat{\theta}_{1}} \right) = \frac{\theta_2}{\theta_1}=\frac{E({\widehat{\theta}_{2}})}{E(\widehat{\theta}_{1})},
\end{equation*}
an interesting result. This is indeed the case for the regression coefficient $b_1$, as seen next.

\noindent (ii)
Let $p=1$ and $n\geq 2 $ so that $E(Y|X)=b_0+b_1X$. Then, as seen earlier,
\begin{eqnarray*}
b_1=\frac{C_{yx}}{C_{xx}}=\frac{\theta_2}{\theta_1} (\mbox{say});~~ \hat{b}_1=\frac{S_{yx}}{S_{xx}},
\end{eqnarray*}
where
$ E(S_{yx})=C_{yx}, E(S_{xx})=C_{xx}.$ Also, by Theorem \ref{thm3.2}
\begin{eqnarray*}
E(\hat{b}_{1})=E(\frac{S_{yx}}{S_{xx}})=\frac{C_{yx}}{C_{xx}}=b_1.
\end{eqnarray*}
The above result holds because the conditions $S_{xx}> 0 ~ a.e. $ and from \eqref{eqn3.16n}
\begin{eqnarray*}
E(\widehat{\theta}_{2}|X)=E(S_{yx}|\mathbf{X})=S_{xx}b_{1}=\widehat{\theta}_1 b_{1}
\end{eqnarray*}
 of Lemma {3.2} are satisfied.}
\end{remark}

\noindent Let $b^{t}=(b_0, b_{(p)})$ denote the row-vector of regression coefficients and $1_{n}$ denote the $n$- dimensional column vector with all its entries equal to $1$. The  least-squares estimator of $b$,
when $\mathbf{X}$ is fixed (called fixed design) is
\begin{equation} \label{eqn3.18}
\widehat{b}_{l}=( \mathbf{X}_{1}^{t}\mathbf{X}_{1})^{-1}\mathbf{X}_{1}^{t} \mathbf{Y},
\end{equation}
where $\mathbf{X}_1=(1_{n} \vdots \mathbf{X})$,  $\mathbf{Y}^t= (Y_1, \ldots, Y_n)$ and $\mathbf{X}$ is defined in \eqref{eqn3.4n}. We now have following result.


\begin{theorem}\label{thm3.3}
Let $p \geq 1$ and  $S_{xx} >0 a.e.$ Then  the unbiased estimator $\widehat{b}_{u}^t=(\widehat{b}_0, \widehat{b}_{(p)})^t$, defined in Theorem \ref{thm3.1}, coincides with the least squares estimator $\widehat{b}_{l}$, defined in \eqref{eqn3.18}, for the fixed design.
\end{theorem}

\noindent {\bf Proof.}
Note first that
\begin{eqnarray*}
\mathbf{X}_{1}^{t} \mathbf{X}_1
=
\begin{pmatrix}
1_{n}^{t} 1_{n} & 1_{n}^{t} \mathbf{X}\\
\mathbf{X}^{t}1_{n} & \mathbf{X}^{t} \mathbf{X}\\
\end{pmatrix}
=
\begin{pmatrix}
n & n \overline{X}_{(p)} \\
n \overline{X}_{(p)}^{t} & \mathbf{X}^{t} \mathbf{X}
\end{pmatrix}
\end{eqnarray*}
Using the formula (see Seber(1984), p.~519) for the inverse of a partitioned matrix, namely,
\vone
$\left(\begin{array}{cc}
A & B \\B^t & D
\end{array}\right)^{-1} = \left(\begin{array}{cc}
A^{-1}+ FG^{-1}F^{t} & -FG^{-1} \\-G^{-1}F^{t} & G^{-1}
\end{array}\right),$
\vone
\noindent where $F=A^{-1}B $ and $G=D-B^{t}F$, we obtain  
\begin{eqnarray} \label{eqn3.19nn}
(\mathbf{X}_{1}^{t} \mathbf{X}_{1})^{-1}=
\begin{pmatrix}
\frac{1}{n}+ \frac{1}{(n-1)} \overline{X}_{(p)} S_{xx}^{-1} \overline{X}_{(p)}^t & \frac{-1}{(n-1)} \overline{X}_{(p)} S_{xx}^{-1} \\
\frac{-1}{(n-1)} S_{xx}^{-1} \overline{X}_{(p)}^{t} & \frac{1}{(n-1)} S_{xx}^{-1}
\end{pmatrix}
\end{eqnarray}
Note for example that
the $(i, j)$- th element of the matrix $ G= (\mathbf{X}^{t} \mathbf{X})- n \overline{X}_{(p)}^{t} \overline{X}_{(p)} $
is
\begin{eqnarray*}
G_{ij}= \sum_{r=1}^{n} X_{ir} X_{jr} -n \overline{X}_{i} \overline{X}_{j}=(n-1) S_{ir}.
\end{eqnarray*}
Hence, $G=(n-1)S_{xx}.$ The other elements of the matrix in the rhs of $\eqref{eqn3.19nn}$ can easily be computed.
Therefore,
\begin{eqnarray*}
(\mathbf{X}_{1}^{t} \mathbf{X}_{1})^{-1} \mathbf{X}_{1}^{t} \mathbf{Y} & =&
\begin{pmatrix}
\frac{1}{n}+ \frac{1}{(n-1)} \overline{X}_{(p)} S_{xx}^{-1} \overline{X}_{(p)}^t & \frac{-1}{(n-1)} \overline{X}_{(p)} S_{xx}^{-1} \\
\frac{-1}{(n-1)} S_{xx}^{-1} \overline{X}_{(p)}^{t} & \frac{1}{(n-1)} S_{xx}^{-1}
\end{pmatrix} \begin{pmatrix}
n & \overline{Y} \\
\mathbf{X}^{t} & Y
\end{pmatrix}\\ \\
&=&
\begin{pmatrix}
\overline{Y}+ \frac{1}{(n-1)} \overline{X}_{(p)} S_{xx}^{-1}(n \overline{X}_{(p)}^t \overline{Y}- \mathbf{X}^{t} Y)  \\
\frac{1}{(n-1)} S_{xx}^{-1} \mathbf{X}^{t} Y - \frac{n}{(n-1)} S_{xx}^{-1} \overline{X}_{(p)}^{t} \overline{Y}
\end{pmatrix}\\ \\
&=&
\begin{pmatrix}
\overline{Y} - \overline{X}_{(p)} S_{xx}^{-1} S_{yx} \\
S_{xx}^{-1} S_{yx}
\end{pmatrix},
\end{eqnarray*}
using the fact $(\mathbf{X}^{t} \mathbf{Y}- n \overline{X}_{(p)} \overline{Y})= (n-1) S_{yx}. $
The result now follows. \hfil{$\Box$}

\vone
\noindent We next find the covariance matrix $ D(\widehat{b}_{u}) =Cov(\widehat{b}_{u}, \widehat{b}_{u})$ of $\widehat{b}_{u} = (\widehat{b}_0, \widehat{b}_{(p)})^t. $

\begin{theorem}\label{thm3.4}
Assume, in the model \eqref{eqn2.1},  $Var(Y|X_{(p)})= \sigma^2_{y|x},$ which does not depend on $X_{(p)}$, and $S_{xx}^{-1}$ exists. Then the covariance matrix of $\widehat{b}_{u}$ is
\begin{eqnarray}
D(\widehat{b}_{u}) &=& \sigma^2_{y|x} E(\mathbf{X}_{1}^{t} \mathbf{X}_{1})^{-1} \nonumber\\
&=& \sigma^2_{y|x} E \begin{pmatrix}
\frac{1}{n}+ \frac{1}{(n-1)} \overline{X}_{(p)} S_{xx}^{-1} \overline{X}_{(p)}^t & \frac{-1}{(n-1)} \overline{X}_{(p)} S_{xx}^{-1} \\
\frac{-1}{(n-1)} S_{xx}^{-1} \overline{X}_{(p)}^{t} & \frac{1}{(n-1)} S_{xx}^{-1} \label{neqn3.18n}
\end{pmatrix}
\end{eqnarray}
\end{theorem}

\noindent {\bf Proof}.
Since $E(\widehat{b}_{u} | \mathbf{X})=b, $ we have
\begin{eqnarray}\label{eqn3.21mm}
D(\widehat{b}_{u}) &=& E(D(\widehat{b}_{u}|\mathbf{X})) \nonumber\\
&=& E \begin{pmatrix}
Var(\widehat{b}_0|\mathbf{X}) & Cov(\widehat{b}_0, \widehat{b}_{(p)}| \mathbf{X}) \\
(Cov(\widehat{b}_0, \widehat{b}_{(p)}| \mathbf{X}))^t & D(\widehat{b}_{(p)}| \mathbf{X})
\end{pmatrix}
\end{eqnarray}

\noindent First we obtain
\begin{eqnarray}\label{eqn3.22n}
D(\widehat{b}_{(p)}| \mathbf{X})
&=& D(S^{-1}_{xx}S_{yx}|\mathbf{X}) \nonumber\\
&=&S^{-1}_{xx}D(S_{yx}|\mathbf{X})S^{-1}_{xx}.
\end{eqnarray}

\noindent Observe now,
\begin{eqnarray}\label{eqn3.18n}
Cov(S_{yr}, S_{ys}|\mathbf{X}) &=& Cov\left(\frac{1}{(n-1)}\sum_{j=1}^n(X_{rj}-\overline{X}_{r})Y_{j},\frac{1}{(n-1)}\sum_{k=1}^n (X_{sk}-\overline{X}_{s})Y_{k}|\mathbf{X}\right)\nonumber\\
&=&\frac{1}{(n-1)^2}\sum_{j,k}(X_{rj}-\overline{X}_{r})(X_{sk}-\overline{X}_{s})Cov((Y_j,Y_k)|\mathbf{X}).
\end{eqnarray}
Since
$Var(Y|X_{(p)})=\sigma^{2}_{y|x}$ , we have
\begin{equation*}
Cov\left((Y_{j},Y_{k})|\mathbf{X}\right)=\left\{ \begin{array}{cl}
\sigma^{2}_{y|x},~\mbox{if}~  k=j\\
0, ~\mbox{otherwise.}
\end{array}\right.
\end{equation*}
\noindent Hence, for $r,s \in \{1,2,\ldots,p\},$
\begin{eqnarray*}
Cov \left((S_{yr},S_{ys})|\mathbf{X}\right) &=& \frac{\sigma^{2}_{y|x}}{(n-1)^2}\sum_{j}(X_{rj}-
\overline{X}_{r})(X_{sj}-\overline{X}_{s})
=\frac{\sigma^{2}_{y|x}}{(n-1)}S_{rs}
\end{eqnarray*}
and
\begin{equation}\label{eqn 3.19}
D(S_{yx}|\mathbf{X})=\frac{\sigma^{2}_{y|x}}{(n-1)}S_{xx}.
\end{equation}
Hence, we obtain from \eqref{eqn3.22n}
\begin{equation} \label{eqn3.25nn}
D(\widehat{b}_{(p)}| \mathbf{X})=\frac{\sigma^{2}_{y|x}}{(n-1)}S^{-1}_{xx}.
\end{equation}
We obtain next $Var(\widehat{b}_{0} | \mathbf{X}).$
Note
\begin{eqnarray} \label{eqn3.26m}
Var(\widehat{b}_{0} | \mathbf{X}) &=&Var(\overline{Y}- \overline{X}_{(p)} \widehat{b}_{(p)}| \mathbf{X}) \nonumber\\
&=& Var(\overline{Y}| \mathbf{X}) + Var(\overline{X}_{(p)} \widehat{b}_{(p)}| \mathbf{X})-2 Cov(\overline{Y}, \overline{X}_{(p)} \widehat{b}_{(p)}| \mathbf{X}).
\end{eqnarray}
By assumption,
\begin{equation}
Var( \overline{Y} | \mathbf{X}) = \frac{\sigma^2_{y|x}}{n}
\end{equation}
and using  $\eqref{eqn3.25nn}$
\begin{eqnarray}\label{eqn3.28m}
Var(\overline{X}_{(p)} \widehat{b}_{(p)} | \mathbf{X})&=& \overline{X} D(\widehat{b}_{(p)}| \mathbf{X}) \overline{X}_{(p)}^{t}
= \frac{\sigma^2_{y|x}}{(n-1)} \overline{X}_{(p)} S_{xx}^{-1} \overline{X}^{t}_{(p)}.
\end{eqnarray}
We next show that
$Cov(\overline{Y}, \overline{X}_{(p)} \widehat{b}_{(p)} | \mathbf{X})=0. $
Since
\begin{equation*}
S_{yk}= \frac{1}{(n-1)} \sum_{j=1}^{n} ( X_{kj}- \overline{X}_{k})Y_{j},
\end{equation*}
we can write
\begin{equation*}
S_{yx}= \frac{1}{(n-1)} \sum_{j=1}^{n} ( X_{(p)j}- \overline{X}_{(p)})^{t} Y_{j},
\end{equation*}
where $X_{(p)j}=(X_{1j}, X_{2j}, \ldots , X_{pj}), $ and $ \overline{X}_{(p)}=(\overline{X}_{1}, \overline{X}_{2}, \ldots, \overline{X}_{p}),$ as before.

\noindent As $Y_{j}$'s are iid, we have
\begin{eqnarray} \label{neqn3.30}
Cov(\overline{Y}, \overline{X}_{(p)} \widehat{b}_{(p)} | \mathbf{X})&=& Cov(\frac{1}{n} \sum_1^n Y_j, \frac{1}{(n-1)} \overline{X}_{(p)} S^{-1}_{xx} \sum_{j=1}^{n} (X_{(p)j} - \overline{X}_{(p)})^{t} Y_j | \mathbf{X}) \nonumber\\
&=& \sum_{j=1}^{n} Cov( \frac{1}{n} Y_{j}, \frac{1}{(n-1)} \overline{X}_{(p)} S^{-1}_{xx}( X_{(p)j}- \overline{X}_{(p)})^{t} | \mathbf{X})\nonumber \\
&=& \sum_{j=1}^{n} \frac{1}{n(n-1)} \overline{X}_{(p)} S^{-1}_{xx}(X_{(p)j}- \overline{X}_{(p)})^{t} Cov(Y_j, Y_j | \mathbf{X}) \nonumber\\
&=& \frac{\sigma^2_{y|x}}{n(n-1)} \overline{X}_{(p)} S^{-1}_{xx} \sum_{j=1}^{n}(X_{(p)j}- \overline{X}_{(p)})^{t}  \nonumber \\
&=& 0.
\end{eqnarray}
Therefore, from $\eqref{eqn3.26m}-\eqref{eqn3.28m},$ we have  
\begin{equation}\label{eqn3.31nn}
Var(\widehat{b}_0| \mathbf{X})= \sigma^{2}_{y|x} \left( \frac{1}{n} + \frac{1}{(n-1)} \overline{X}_{(p)} S^{-1}_{xx} \overline{X}_{(p)}^{t}\right).
\end{equation}
Finally, we compute
\begin{equation}
Cov(\widehat{b}_0, \widehat{b}_{(p)}| \mathbf{X})= \left(Cov( \widehat{b}_{0}, e_{1}^{t}  \widehat{b}_{(p)}| \mathbf{X}), \ldots, Cov(\widehat{b}_0, e^{t}_{p} \widehat{b}_{p}| \mathbf{X})\right),
\end{equation}
where $e_{j}^{t}$ is the $(1 \times p)$ row vector whose j-th entry is 1 and all other entries are zeros.\\
Note, for $ 1 \leq r \leq p, $
\begin{equation}
Cov(\widehat{b}_0, e^{t}_{r}\widehat{b}_{(p)}| \mathbf{X})=Cov(\overline{Y}, e^{t}_{r}\widehat{b}_{(p)}| \mathbf{X})-Cov(\overline{X}_{(p)} \widehat{b}_{(p)},  e^{t}_{r}\widehat{b}_{(p)}| \mathbf{X}).
\end{equation}
It can be shown, as in \eqref{neqn3.30}, that
\begin{equation}
Cov(\overline{Y}, e^{t}_{r} \widehat{b}_{(p)}| \mathbf{X})=0.
\end{equation}
Also,  for $ 1 \leq r \leq p, $\begin{eqnarray}
Cov(\overline{X}_{(p)} \widehat{b}_{(p)}, e^{t}_{j} \widehat{b}_{(p)}| \mathbf{X}) &=& \overline{X}_{(p)} D(\widehat{b}_{(p)}|\mathbf{X}) e_{r}
=\frac{\sigma^2_{y|x}}{(n-1)} \overline{X}_{(p)} S^{-1}_{xx} e_{r}.
\end{eqnarray}
Hence,
\begin{eqnarray}\label{eqn3.36nn}
Cov(\widehat{b}_0, \widehat{b}_{(p)}| \mathbf{X})&=& -\frac{\sigma^2_{y|x}}{(n-1)} \overline{X}_{(p)}S^{-1}_{xx}(e_1,e_2,\ldots,e_p) =
- \frac{\sigma^{2}_{y|x}}{(n-1)}\overline{X}_{(p)} S^{-1}_{xx}.
\end{eqnarray}
Substituting $ \eqref{eqn3.31nn},\eqref{eqn3.36nn}$ and $\eqref{eqn3.25nn}$ in $\eqref{eqn3.21mm}$ and using $\eqref{eqn3.19nn}$, the result follows. \hfill{$\Box$}

\subsection{Analysis of Residuals and Regression Sum of Squares}
Consider the model \eqref{eqn3.5n} defined by 
\begin{equation} \label{neqn3.34}
 E(Y_j|X_{(p)j}) =b_0 + X_{(p)j}b_{(p)}, ~ Var(Y_j|X_{(p)j})=\sigma_{y|x}^2,  ~ 1 \leq j \leq n.
\end{equation}
Assume $n>(p+1)$, and let $b^{t}= (b_0, b_{(p)})$, as before. Then the above model can be written as
\begin{equation}\label{cneun1}
 E(\mathbf{Y}|\mathbf{X_{1}})=\mathbf{X_{1}} b;\;\; D(\mathbf{Y}|\mathbf{X_{1}})=\sigma_{y|x}^{2}I_{n},
\end{equation}
where $\mathbf{Y}^t=(Y_1, \ldots, Y_n)$, $\mathbf{X_{1}}= (1_n \vdots  \mathbf{X})$ and $I_n$ is the identity matrix of order $n$.

\noindent Then, in view of Theorem \ref{thm3.3}, 
\begin{equation*}
 \widehat{b}_u=(\mathbf{X}_{1}^{t}\mathbf{X_{1}})^{-1}\mathbf{X}_{1}^{t}\mathbf{Y}.
\end{equation*}

\noindent Let $e=(\mathbf{Y}-\widehat{\mathbf{Y}})$, where $\widehat{\mathbf{Y}}={\mathbf{X}}_{1}\widehat{b}_u$, denote the residual vector. Then 
\begin{equation*}
 e=\big[I_{n}-\mathbf{X}_{1}(\mathbf{X}_{1}^{t}\mathbf{X_{1}})^{-1}\mathbf{X}_{1}^{t}\big] \mathbf{Y}= M \mathbf{Y} \;\; (\mathrm{say}),
\end{equation*}
where $M= M(\mathbf{X})=\big[I_{n}-\mathbf{X}_{1}(\mathbf{X}_{1}^{t}\mathbf{X_{1}})^{-1}\mathbf{X}_{1}^{t}\big]$ is symmetric and idempotent. 
It is also well known (see Draper and Smith (2002)) that $tr(M)=(n-p-1)$ and $M\mathbf{X}_{1}=0$ a.s.\\
Note also that $e^{t}e=\mathbf{Y}^{t}M\mathbf{Y}$ and hence the  the mean residual sum of squares is
\begin{eqnarray}\label{cneqn2}
 MSS_{E}&=&E(e^{t}e)\nonumber\\
 &=&E\left\{E(\mathbf{Y}^{t}M\mathbf{Y}|\mathbf{X}_{1})\right\}\nonumber\\
 &=&E\left\{(E(\mathbf{Y}|\mathbf{X}_{1}))^{t}M(E(\mathbf{Y}|\mathbf{X}_{1}))+tr (MD(\mathbf{Y}|\mathbf{X}_{1}))\right\}\nonumber\\
 &=&E\left\{b^{t}{\mathbf{X}_{1}}^{t}M\mathbf{X}_{1}b +tr(M)\sigma_{y|x}^{2}I_{n}\right\}\nonumber\\
 &=&\sigma_{y|x}^{2}tr(M)\nonumber\\
 &=& (n-p-1)\sigma_{y|x}^{2}.
\end{eqnarray}
Thus, an unbiased estimator of $\sigma_{y|x}^{2}$ is 
\begin{equation}\label{cneqn3}
 \widehat{\sigma^2}_{y|x}=\frac{SS_{E}}{(n-p-1)}=\frac{e^{t}e}{(n-p-1)}.
\end{equation}
Indeed, we have (see \eqref{cneqn2}) shown that
\begin{equation}\label{cneqn3n}
 E(\widehat{\sigma^2}_{y|x}|\mathbf{X}_{1}) =\sigma_{y|x}^{2}
\end{equation}
so that it is also conditionally unbiased.

\begin{theorem} {\em
Let  $D(\widehat{b}_{u})$, given in  \eqref{neqn3.18n}, be the dispersion matrix unbiased estimator $\widehat{b}_{u}$. The its unbiased estimator is given by
\begin{equation}
\widehat{D} (\widehat{b}_{u})= \widehat{\sigma^2}_{y|x}(\mathbf{X}_{1}^{t} \mathbf{X}_{1})^{-1},
\end{equation}
where $\widehat{\sigma^2}_{y|x} $ is defined in \eqref{cneqn3}.}
\end{theorem}

\noindent {\bf Proof}.
Using \eqref{cneqn3n}, we get
\begin{align*}
E(\widehat{D} (\widehat{b}_{u}))=& E( E(\widehat{\sigma^2}_{y|x}(\mathbf{X}_{1}^{t} \mathbf{X}_{1})^{-1}|\mathbf{X}_{1}))\\
 =& E(E(\widehat{\sigma^2}_{y|x}|\mathbf{X}_{1})(\mathbf{X}_{1}^{t} \mathbf{X}_{1})^{-1})\\
 =& E({\sigma_{y|x}^{2}}(\mathbf{X}_{1}^{t} \mathbf{X}_{1})^{-1})\\
 =&    {\sigma_{y|x}^{2}} E((\mathbf{X}_{1}^{t} \mathbf{X}_{1})^{-1})\\
=& D(\widehat{b}_{u})
\end{align*}
and hence the result follows.

\subsection{Mean Regression Sum of Squares}
\noindent Note first that
\begin{equation*}
 \mathbf{X}_{1}^{t}e=\mathbf{X}_{1}^{t}M \mathbf{Y}=(M\mathbf{X}_{1})^{t}\mathbf{Y}= \mathbf{0}\;\;\Longrightarrow {\widehat{\mathbf{Y}}}^{t}e={\bf 0}~ a.s.
\end{equation*}
In particular, we have ${\bf1}_{n}^{t}e=0$~~~$\Longrightarrow \overline{Y}=\overline{\widehat Y} a.s.$\\
In addition,
\begin{eqnarray}\label{cneqn4}
 Cov(\widehat b_u, e|\mathbf{X}_{1})&=& Cov((\mathbf{X}_{1}^{t}\mathbf{X_{1}})^{-1}\mathbf{X}_{1}^{t}\mathbf{Y}, \mathbf{Y}|\mathbf{X}_{1})
 -Cov(\widehat b_u, X\widehat b_u |\mathbf{X}_{1})\nonumber\\
 &=& (\mathbf{X}_{1}^{t}\mathbf{X_{1}})^{-1}\mathbf{X}_{1}^{t}D(\mathbf{Y}|\mathbf{X_{1}})-D(\widehat b_u|\mathbf{X}_{1})\mathbf{X}_{1}^{t}\nonumber\\
 &=& \sigma_{y|x}^{2}(\mathbf{X}_{1}^{t}\mathbf{X_{1}})^{-1}\mathbf{X}_{1}^{t}
 -\sigma_{y|x}^{2}(\mathbf{X}_{1}^{t}\mathbf{X_{1}})^{-1}\mathbf{X}_{1}^{t}\nonumber\\
 &=& {\bf 0}.
\end{eqnarray}

\noindent Hence,
\begin{equation*}
 Cov(\widehat b_u,e)=E\big(Cov(\widehat b_u, e|\mathbf{X_{1}}\big)+Cov\big(E(\widehat{b}_u|\mathbf{X_{1}}),E(e|\mathbf{X_{1}})\big)={\bf 0},
\end{equation*}
using (\ref{cneqn4}) and $E(e|\mathbf{X_{1}})=\mathbf{X_{1}}b-\mathbf{X_{1}}E(\widehat{b}_u|\mathbf{X_{1}})={\bf 0}$ (see \eqref{eqn3.16n} and \eqref{eqn3.18}).\\
Note (\ref{cneqn4}) also implies $Cov(\widehat Y,e|\mathbf{X_{1}})={\bf 0}.$ Also,
\begin{equation*}
 {\mathbf{Y}^{t}}\mathbf{Y}=(e+\widehat{\mathbf{Y}})^{t}(e+\widehat{\mathbf{Y}})={\widehat{\mathbf{Y}}}^{t}\widehat{\mathbf{Y}}+e^{t}e,
\end{equation*}
which implies
\begin{eqnarray}\label{Correction1}
 E({\mathbf{Y}}^{t}\mathbf{Y}- n\bar{Y}^2)&=&E(\widehat{\mathbf{Y}}^{t}{\widehat{\mathbf{Y}}}-n\bar{Y}^2)+E(e^{t}e).
 \end{eqnarray}
That is,
 \begin{eqnarray*}
 MSS_{T}&=&MSS_{R}+MSS_{E},
\end{eqnarray*}
where $MSS_{T}$ and $MSS_{R}$ respectively denote the mean total sum of squares and mean regression sum of squares.\\
The mean regression sum of squares $MSS_{R}$ is given by
\begin{eqnarray}\label{cneqn5}
 E({\widehat{\mathbf{Y}}}^{t}{\widehat{\mathbf{Y}}})&=&E\left\{E({\widehat b_u}^{t}\mathbf{X_{1}}^{t}\mathbf{X_{1}}{\widehat b_u}|\mathbf{X_{1}})\right\}\nonumber\\
 &=&E\left\{(E(\widehat b_u|\mathbf{X_{1}}))^{t}\mathbf{X_{1}}^{t}\mathbf{X_{1}}E(\widehat b_u|\mathbf{X_{1}})
 +tr(\mathbf{X_{1}}^{t}\mathbf{X_{1}})D(\widehat b_u|\mathbf{X_{1}})\right\}\nonumber\\
 &=&E\left\{b^{t}\mathbf{X_{1}}^{t}\mathbf{X_{1}} b+\sigma_{y|x}^{2}tr(I_{p+1})\right\}\nonumber\\
 &=&E\left\{b^{t}\mathbf{X_{1}}^{t}\mathbf{X_{1}} b+(p+1)\sigma_{y|x}^{2}\right\}.
\end{eqnarray}
Thus, we obtain
\begin{equation}\label{cneqn6}
 E\bigg(\frac{{\widehat{\mathbf{Y}}}^{t} {\widehat{\mathbf{Y}}}}{p+1}\bigg)= \frac{1}{(p+1)}E(b^{t}(\mathbf{X_{1}}^{t}\mathbf{X_{1}})b)+\sigma_{y|x}^{2}.
\end{equation}
Form (\ref{Correction1}), (\ref{cneqn2}) and (\ref{cneqn5}), we obtain the mean total sum of squares as 
\begin{equation*}
 E({\mathbf{Y}}^{t}\mathbf{Y})=E(b^{t}(\mathbf{X_{1}}^{t}\mathbf{X_{1}}) b)+n \sigma_{y|x}^{2}
\end{equation*}
leading to 
\begin{equation*}
  E(\frac{{\mathbf{Y}}^{t}\mathbf{Y}}{n})=\frac{1}{n}E(b^{t}(\mathbf{X_{1}}^{t}\mathbf{X_{1}}) b)+\sigma_{y|x}^{2}.
\end{equation*}
Finally, we define the mean coefficient of determination as 
\begin{equation*}
 \mathbb{R}_{m}^{2}=E(R_{0}^{2})=E\bigg(\frac{\sum_{1}^{n}(\widehat Y_{i}-\overline{Y})^{2}}{\sum_{1}^{n}( Y_{i}-\overline{Y})^{2}}\bigg),
\end{equation*}
so that $R_{0}^{2}$ is an unbiased estimator of $\mathbb{R}_{m}^{2}$. Note $\mathbb{R}_{m}$
denote the mean multiple correlation coefficient.

\vone
\noindent Finally, consider the problem of prediction of $E(Y_0)$ or that of $Y_0$, 
a future observation corresponding to $X_{(p)0}$, the new observed vector of predictors can easily be considered. That is,
$\widehat{Y}_0=X_{(p)0}\widehat{b}_{(p)}$
may be considered as a predictor of $E(Y_0)$ or that of $Y_0$.
In that case, we have
\begin{eqnarray*}
Var(\widehat{Y}_0)&=& X_{(p)0}D(\widehat{b}_{(p)})X^{t}_{(p)0}
=\frac{{\sigma^{2}_{y|x}}}{(n-1)}X_{(p)0}E(S^{-1}_{xx})X^{t}_{(p)0}.
\end{eqnarray*}
Also, an estimated value of $Var(\widehat{Y}_0)$ is
\begin{center}
$\widehat{V}ar(\widehat{Y}_0)=\displaystyle \frac{{\widehat{\sigma}^{2}_{y|x}}}{(n-1)}X_{(p)0}S^{-1}_{xx}X^{t}_{(p)0},$
\end{center}
which could be used to obtain prediction intervals for $Y_0$ or that of $E(Y_0)$.

\section{Application to $AR(p)$ Process}
In this section, we show that the unbiasedness approach works for $AR(p)$ time series models also. 
Consider $AR(p)$ process so that
\begin{equation} \label{eqn0}
Y_{t}=\phi_{0}+\phi_{1}Y_{t-1}+\ldots+\phi_{p}Y_{t-p}+
\varepsilon_{t},
\end{equation}
where $\varepsilon_{t}$'s are i.i.d with $E(\varepsilon_{t})=0$ and
$V(\varepsilon_{t})=\sigma^{2}$.
Note  $AR(p)$ process  satisfies
\begin{equation}\label{eqn12}
E(Y_{t}|Y_{t-1},\ldots,Y_{t-p})=\phi_{0}+\phi_{1}Y_{t-1}+\ldots+\phi_{p}Y_{t-p}, ~~
\end{equation}
which is of the form \eqref{eqn2.1}.
Let $ E(Y_{t})=\eta ,$ and from model (\ref{eqn12}),
\begin{equation*}
C_{yx} =
\begin{pmatrix}
Cov(Y_{t},Y_{t-1})\\
Cov(Y_{t},Y_{t-2})\\
\vdots\\
 Cov(Y_{t},Y_{t-p})
\end{pmatrix}
=
\begin{pmatrix}
\gamma(1)\\
\gamma(2) \\
\vdots \\
\gamma(p)\\
\end{pmatrix} ~~,
\end{equation*}
and

\begin{eqnarray*}
C_{xx}=(Cov(Y_{t-i},Y_{t-j}))
&=&(\gamma_{(j-i)}), ~~
\end{eqnarray*}  
for $1\leq i\leq j\leq p$ ~~.\\
Then from Theorem \ref{thm2.1}, we have
\begin{equation}\label{eqn4}
\begin{pmatrix}
\gamma(0) & \ldots& \gamma(p-1)\\
\gamma(1) &\ldots & \gamma(p-2)\\
& \vdots& \\
\gamma(p-1) & \ldots & \gamma(0)
\end{pmatrix}
\begin{pmatrix}
\phi_{1}\\
\phi_{2}\\
\vdots\\
\phi_{p}
\end{pmatrix}
=
\begin{pmatrix}
\gamma(1)\\
\gamma(2) \\
\vdots \\
\gamma(p)\\
\end{pmatrix}
\end{equation}
or equivalently dividing by $ \gamma(0)= Var(Y_{t}), $

\begin{equation}\label{eqn5}
\begin{pmatrix}
1 & \rho(1) &\ldots& \rho(p-1)&\\
\rho(1) & 1 &\ldots & \rho(p-2)&\\
& \vdots& \\
\rho(p-1) & \rho(p-2) &\ldots & 1&
\end{pmatrix}
\begin{pmatrix}
\phi_{1}\\
\phi_{2}\\
\vdots\\
\phi_{p}
\end{pmatrix}
=
\begin{pmatrix}
\rho(1)\\
\rho(2) \\
\vdots \\
\rho(p)\\
\end{pmatrix}
\end{equation}
and
\begin{eqnarray}\label{eqn6}
\phi_{0}= \eta-(\eta, \ldots,\eta) \begin{pmatrix}
\phi_{1}\\
\phi_{2}\\
\vdots\\
\phi_{p}
\end{pmatrix}
=\eta(1-\sum_{j=1}^{p}{\phi_{j}}).
\end{eqnarray}
Note also that \eqref{eqn5} is nothing but
Yule-Walker equations.

\noindent Assume we have a sample
$(Y_{1}, Y_{2}, \ldots  , Y_{n}) ,~~n \geq p+1 $, that satisfies
\eqref{eqn0}. Then,
\begin{eqnarray}\label{eqn2}
Y_{p+1}&=&\phi_{o}+\phi_{1}Y_{p}+\ldots+\phi_{p}Y_{1}+
\varepsilon_{p+1}\nonumber \\
Y_{p+2}&=&\phi_{o}+\phi_{p+1}Y_{p+1}+\ldots+\phi_{p}Y_{2}+
\varepsilon_{p+2}\nonumber \\
&\vdots& \\
Y_{n}&=&\phi_{o}+\phi_{1}Y_{n-1}+\ldots+\phi_{p}Y_{n-p}+
\varepsilon_{n}. ~~\nonumber
\end{eqnarray}
Alternatively, the equations in  \eqref{eqn12} can also be written as,  for $ 1\leq j \leq n-p$,
\begin{equation}\label{eqn31}
E(Y_{p+j}|Y_{p+j-1,\ldots,Y_{j}})=\phi_{o}+\phi_{1}Y_{p+j-1}+\ldots+\phi_{p}Y_{j}.~~
\end{equation}
Let now \begin{equation*} \mathbf{Y}=
\begin{pmatrix}
Y_{p+1} \\
Y_{p+2} \\
\vdots \\
Y_{n}
\end{pmatrix},
     ~~~~~~ \phi=
\begin{pmatrix}
\phi_{1} \\
\phi_{2} \\
\vdots \\
\phi_{p} ~~,\\
\end{pmatrix} ~~;
~~ X=
\begin{pmatrix}
Y_{p} & Y_{p-1} &\ldots& Y_{1} & \\
Y_{p+1} & Y_{p} &\ldots& Y_{2}  &\\
& \vdots & \\
Y_{n-1} & Y_{n-2} &\ldots& Y_{n-p} & \\
\end{pmatrix}
\end{equation*}
Let us introduce the following notations.
Define for $ 0 \leq i \leq p $ , \begin{equation*}
\overline{Y}_{p-i+1} = \frac{1}{(n-p)} \sum^{n-p}_{k=1}{Y_{p-i+k}},
\end{equation*}
the mean of values in $i$-th column of X.
Also, let for $1\leq i, j \leq p $,
\begin{equation*}
S_{i,j}=
\frac{1}{(n-p-1)}\sum^{n-p}_{k=1}{(Y_{p-i+k}-\overline{Y}_{p-i+1})(Y_{p-j+k}-\overline{Y}_{p-j+1})}
\end{equation*}
be the sample covariance between the $ i$-th and $j$-th columns
of X.\\
Similarly,  for $1\leq j \leq p$, let
\begin{equation*}
S_{p+1,j}=\frac{1}{(n-p-1)}\sum^{n-p}_{l=1}{(Y_{p+l}-\overline{Y}_{p+1})(Y_{p-j+l}-\overline{Y}_{p-j+1})}
\end{equation*}
be the sample covariance between Y and $j$-th column of X.
Then
\begin{equation*}
S_{y,x}=
\begin{pmatrix}
S_{p+1,1}\\
S_{p+1,2} \\
\vdots \\
S_{p+1,p}
\end{pmatrix}
;~~S_{xx}=
\begin{pmatrix}
S_{1,1} & S_{1,2} &\ldots& S_{1,p}&\\
S_{2,1} & S_{2,2} &\ldots & S_{2,p} &\\
&\vdots& \\
S_{p,1} & S_{p,2} &\ldots& S_{p,p} & \\
\end{pmatrix}.
\end{equation*}

\noindent Using our formulas in Theorem \ref{thm3.1} for estimation of $\phi$ , we
get,
\begin{equation}
S_{xx} \widehat{\phi} = S_{yx}.
\end{equation}
That is, $\hat{\phi}_{j}, ~1 \leq j \leq p$, satisfies
\begin{eqnarray} \label{neqn410}
\begin{pmatrix}
S_{1,1} & S_{1,2} &\ldots& S_{1,p}&\\
S_{2,1} & S_{2,2} &\ldots & S_{2,p} &\\
&\vdots& \\
S_{p,1} & S_{p,2} &\ldots& S_{p,p} & \\
\end{pmatrix}
\begin{pmatrix}
\hat{\phi_{1}}\\
\hat{\phi_{2}}\\
\vdots\\
\hat{\phi_{p}}
\end{pmatrix}
=
\begin{pmatrix}
S_{p+1,1}\\
S_{p+1,2} \\
\vdots \\
S_{p+1,p}\\
\end{pmatrix}
\end{eqnarray}

\noindent Let now $\overline{Y}_{(p)}^{*}=(\overline{Y}_{p},\overline{Y}_{p-1}, \ldots , \overline{Y}_{1})$,
so that we obtain from
(\ref{eqn6}),
\begin{eqnarray}\label{eqn7}
\hat{\phi}_{0}=\overline{Y}_{p+1}-\overline{Y}_{(p)}^{*}\begin{pmatrix}
\hat{\phi}_{1} \\
\hat{\phi}_{2} \\
\vdots \\
\hat{\phi}_{p} ~~,\\
\end{pmatrix}
=\overline{Y}_{p+1}- \sum_{j=1}^{p}(\overline{Y}_{j}~\hat{\phi}_{p+1-j}).
\end{eqnarray}

\noindent Then the fitted $AR(p)$ model is
\begin{equation}
Y_{t}= \hat{\phi_{0}} +
\hat{\phi_{1}}Y_{t-1} +\cdots +\hat{\phi_{p}}Y_{t-p},
\end{equation}
where $\hat{\phi_{j}}, 0 \leq j \leq p, $ satisfy  (\ref{neqn410}) and (\ref{eqn7}).

\begin{remark} {\em Using our method, we have given a rigorous proof for the unbiased estimators of the parameters of the $AR(p)$ model. This is another advantage of our approach.}
\end{remark}

\subsection{Numerical examples}
\noindent Finally, we discuss in this section two numerical examples, one based on simulated data and the other based on real-life data, for estimating the parameters of the $AR(3)$ model. We show our results yield the estimates that coincide with classical estimates.

\vspace*{-0.1cm}
\begin{example} {\em We simulated n=100 data from $AR(3)$ model with the parameters (0.4, 0.1, 0.3), using the $R$ package. Based on the simulated data, we fitted the $AR(3)$ model and
 estimated regression coefficients, including the intercept term, using our results. 
The results are given below:
\vtwo
\begin{tabular}{|c|c| c| c|}
\hline
Parameters & Least-squares  & Yule-Walker & Unbiased  \\
           & estimate & estimate & estimate\\
\hline
$\phi_{0}$ & 0.0208196 & - & 0.0208196 \\
\hline
$\phi_{1}$ & 0.2875461 & 0.2849 & 0.287546 \\
\hline
$\phi_{2}$ & 0.0935658  & 0.0941 & 0.0935658  \\
\hline
$\phi_{3}$ & 0.3889368  & 0.3542 & 0.3889368 \\
\hline
\end{tabular} }
\end{example}

\newpage
\begin{example} {\em The data in feet on ``Level of lake Huron  data "  (Brockwell and Davis (2006), p.~555)) is fitted for $AR(3)$ model. The results are given below .
\vtwo
\begin{tabular}{|c|c| c| c|}
\hline
Parameters & Least-squares  & Yule-Walker & Unbiased  \\
           & estimate & estimate & estimate\\
\hline
$\phi_{0}$& 1.6460378 & - & 1.6460378 \\
\hline
$\phi_{1}$& 1.0719382 & 1.088704 & 1.0719382 \\
\hline
$\phi_{2}$ & -0.365349  & -0.404544 & -0.365349 \\
\hline
$\phi_{3}$ & 0.1087551  & 0.130754 & 0.1087551 \\
\hline
\end{tabular}}
\end{example}

\vone
\noindent From both the examples above for  the $AR(3)$ model, we see that the unbiased  estimates obtained using our results  coincide with the least-squares and the Yule-Walker estimates, as expected.

\vspace*{.4cm}
\noindent {\bf Acknowledgments}. The author is thankful to Saumya Bansal, Neha Rani Gupta and Swati Verma for some useful discussions and the computational work.

\vspace*{.6cm}
\noindent{\bf{\Large References}}

\vspace*{.3cm}
\noindent
Brockwell, P. J. and Davis, R. A. (2006). {\it Time Series: Theory and Methods}. Second Edition, Springer,
New York.\\
Chatterji, S. and Hadi, A. S. (2006). {\it Regression Analysis by Examples}. John Wiley \& Sons, New Jersey.\\
Draper, N and Smith, H. (2003). {\it Applied Regression Analysis.} Third Edition, Wiley, New York.\\
Quenonuille, M. H. (1950). An application of least squares to family diet surveys. {\it Econometrica}, 18,27-44.\\
Yan, X. and Su, X. G. (2009). {\it Linear Regression Analysis: Theory and Computing}. World Scientific Publishing Co, Pte Ltd, Singapore.
\\


\end{document}